\documentclass[copyright,creativecommons]{eptcs}
\providecommand{\event}{AMCA-POP 2010} 

\providecommand{\firstpage}{1}

\usepackage{breakurl}             
\usepackage{verbatim}
\usepackage{cite}
\usepackage{textcomp}
\usepackage{amssymb}
\usepackage{marvosym}
\usepackage{amsmath}
\usepackage{latexsym}
\usepackage{amsthm}
\usepackage{pstricks,pst-node,pst-text,pst-3d}

\usepackage{graphicx}

\usepackage{comment}

\newcommand{\qqop}[1]{\mathrel{\makebox[2em]{$#1$}}}

\newcommand{\agr}{\quad\big|\quad}
\newcommand{\agrette}{\ \ \ \big|\ \ \ }

\newcommand{\EE}{\mathcal{E}}

\newcommand{\RR}{\mathcal{R}}
\newcommand{\PP}{\mathcal{P}}

\newcommand{\Seq}{\mathcal{S}}
\newcommand{\NPar}{\mathcal{C}}
\newcommand{\EI}{\mathcal{I}}

\newcommand{\TT}{\mathcal{T}}

\newcommand{\Loop}[1]{\ensuremath{\big(#1\big)^L}}
\newcommand{\smallLoop}[1]{\ensuremath{(#1)^L}}

\newcommand{\widesrewrites}[1]{\; \stackrel{#1}{\longmapsto} \;}

\newcommand{\into}{\ensuremath{\;\rfloor\;}}
\newcommand{\pipe}{\ensuremath{\;|\;}}

\newcommand{\extlist}{E}
\newcommand{\envinf}{I}

\theoremstyle{plain}
\newtheorem{definition}{Definition}

\theoremstyle{definition}

\newtheorem{algo}{Algorithm}

\title{Modelling the Dynamics of an {\it Aedes albopictus} Population}
\author{Thomas Anung Basuki \qquad\qquad Antonio Cerone
\institute{International Institute for Software Technology (UNU-IIST)\\
United Nations University, 
Macau SAR China}
\email{\quad anung@iist.unu.edu \quad\qquad antonio@iist.unu.edu}
\and
Roberto Barbuti \quad\qquad Andrea Maggiolo-Schettini \quad\qquad Paolo Milazzo
\institute{Dipartimento di Informatica\\
Universit\`a di Pisa, 
Pisa, Italy}
\email{\quad barbuti@di.unipi.it \quad\qquad maggiolo@di.unipi.it \quad\qquad milazzo@di.unipi.it}
\and
Elisabetta Rossi
\institute{Dipartimento di Coltivazione e Difesa delle Specie Legnose\\
Universit\`a di Pisa, 
Pisa, Italy}
\email{erossi@agr.unipi.it}
}
\def\titlerunning{Modelling the Dynamics of an {\it Aedes albopictus} Population}
\def\authorrunning{T.A. Basuki et al.}
\begin{document}
\maketitle

\begin{abstract}
We present a methodology for modelling population dynamics with formal
means of computer science. This allows unambiguous description of systems and
application of analysis tools such as simulators and model checkers.
In particular, the dynamics of a population of {\it Aedes albopictus} (a species of
mosquito) and its modelling with the Stochastic Calculus of Looping Sequences
(Stochastic CLS) are considered.
The use of Stochastic CLS to model population dynamics requires an extension
which allows environmental events (such as changes in the temperature and rainfalls)
to be taken into account.
A simulator for the constructed model is developed via translation into
the specification language Maude, and used to compare the dynamics obtained from
the model with real data.
\end{abstract}

\section{Introduction}

In the last few years many formalisms have been defined to model biological
systems at molecular and cellular levels
\cite{barcarmagmilpar08,Biocham06,danoslaneve04,priqua05,regevetal04}.
These formalisms allow
unambiguous description of systems and application of analysis tools, such as
simulators and model checkers. 

Among these formalisms the Calculus of Looping Sequences (CLS)
\cite{barcarmagmilpar08} seems to
be applicable to other classes of biological systems.
CLS is based on term rewriting, in which terms may represent
simple biological structures and
compartments, and rewrite rules may represent very general events.
Moreover, a
stochastic extension of CLS has been defined, called Stochastic CLS
\cite{barmagmiltibtro08},
which allows the dynamics over time of the described system to be studied
\cite{cellcycleSpCLS,SCLSMaude}.

In this paper we deal with the problem of modelling population dynamics with
formal means of computer science. Many aspects of population dynamics, such as
births, deaths and interaction of individuals, can be modelled by using
Stochastic CLS. Other aspects related to environmental events, such as changes
in climatic conditions, require an extension of the formalism. In this paper we
define such an extension and use it to model the dynamics of a population of
{\em Aedes albopictus}.

{\em Aedes albopictus} (Skuse), or Asian tiger mosquito, is a species
indigenous to the oriental region, but it is now widespread in many countries
throughout the world. It is an aggressive mosquito, which causes nuisance and
it is well known as an important disease vector \cite{mhskm03}.

A simulator for the constructed model is developed via translation into
the specification language Maude \cite{Maude2:03}, and used to compare the
dynamics obtained from the model with real data.

There are a number of other approaches to the modelling of population
dynamics with formal means of computer science \cite{BMMT09,ccmpps10,mns08}.
Barbuti \emph{et al.} \cite{BMMT09} extend P systems with features typical
of timed automata with the aim of describing periodic environmental events such as
changes of seasons.
Cardona \emph{et al.} \cite{ccmpps10} propose a modelling framework based on P systems
and apply it to the modelling of the dynamics of some scavenger birds in
the Pyrenees.
McCaig, Norman and Shankland \cite{mns08} present a process algebraic approach
to the modelling of population dynamics.
With respect to these proposals we believe that our approach allows a
finer modelling of environmental events. Moreover, thanks to the extensions of
the tools already developed for Stochastic CLS, it offers means for accurate
analysis of phenomena.

\section{Stochastic CLS}\label{sect:SCLS}

Calculi of Looping Sequences (CLS class) is a class of formalisms introduced in
Milazzo's PhD thesis~\cite{milazzo} for modelling biological systems.
The first formalism of the class to be defined, the Full Calculus of
Looping Sequences (Full CLS) uses 4 operators: sequencing, parallel composition,
looping and containment.
The parallel composition operator has the typical semantics as in other formalisms
such as the $\pi$-Calculus and Brane Calculi.
Sequencing is inspired by the sequential structure of several macromolecules such
as DNA.
The looping operator is always applied together with the containment operator
and supports the modelling of membrane-like structures. 
An important language of the CLS class is Stochastic CLS, which supports the
modelling of quantitative aspects of biological systems such as time and
probabilities.

We start by introducing the syntax of sequences and terms, the basic building blocks
of Stochastic CLS.

\begin{definition} Sequences $S$ and Terms $T$ are defined as follows:
\begin{align*}
 S \qqop{::=} \epsilon \agr S\cdot S \agr a \qquad \qquad \qquad
 T \qqop{::=} S \agr \Loop{T} \into T \agr T \pipe T
\end{align*}
where $\epsilon$ represents the empty sequence and $a \in \EE$.
We denote the set of all terms with $\TT$, and the set of all sequences with $\Seq$.
\end{definition}
We assume the existence of a possibly infinite set of symbols $\EE$.
The parallel composition operator $\pipe$ is used to model a mixture of
elements.
The application of the looping and containment operator to two terms $T_1$ and
$T_2$,
denoted by $\Loop{T_1} \into T_2$, models structure $T_2$ within a
compartment surrounded by structure $T_1$.
Structure $T_1$ is called the \emph{loop part} and $T_2$ is called its
\emph{content part}. 

The behaviour of a biological system is modeled by means of transitions
between terms. This is done by applying \emph{rewrite rules}, that are described by
two \emph{patterns}, to be instantiated by terms, and a \emph{rate} that defines
the frequency with which the rule is applied.

\begin{definition}\label{def:pattern} 
Let $TV$ be an infinite set of term
variables ranged over by $X,Y,Z,\ldots$
Term Patterns $TP$ and Patterns $P$ are defined as follows:
\begin{align*}
 TP \qqop{::=} S \agr \Loop{P} \into P \agr TP \pipe TP \qquad \qquad \qquad
 P \qqop{::=} TP \agr TP \pipe X
\end{align*}
where $X \in TV$. We denote with $\PP$ the set of all patterns.
We denote with $Var(P)$ the set of variables in $P$.
\end{definition}

\begin{definition}\label{def:instantiation}
An \emph{instantiation} is a partial function $\sigma:TV\rightarrow \TT$.
We denote with $\Sigma$ the set of all possible instantiations.
Given $P \in \PP$, we denote with $P\sigma$ the term obtained by replacing
all variables $X \in Var(P)$ with $\sigma(X)$.
\end{definition}

\begin{definition} A \emph{rewrite rule} is a triple 
$(P_L,k,P_R)$, denoted with $P_L \! \stackrel{k}{\mapsto} \! P_R$,
where $P_L,P_R \in \PP$, $k \in\mathbb{R}$
and such that $Var(P_R) \subseteq Var(P_L)$.
\end{definition}

\begin{definition}
A biological system is a pair $(T,\RR)$, where $T$ is a term
representing the initial state of the system and $\RR$ is a set of rewrite
rules that represent the potential \emph{events} which may occur in the system.
\end{definition}

Interactions between populations in a biological or ecological system occur
through some kind of reactions, which may be biochemical reactions at molecular
level or changes in organisms' development in ecosystems.
To perform in silico analysis of a biological or ecological system, the behaviour
of the system must be simulated (\emph{in silico experiment}).
The problem of simulating chemically reacting system was stated by
Gillespie~\cite{gillespie77}.
We generalise Gillespie's formulation of the problem to any biological or ecological
system as follows.
\begin{quote}
A volume or environment \emph{V} contains a mixture of \emph{N} species
{$S_1,\ldots,S_N$} which
can interreact through \emph{M} reaction channels ($R_1,\ldots,R_M$).
Given the initial numbers of individuals (molecules or organisms) of each species,
what will these population levels be at any later time?
\end{quote}
Gillespie consider time evolution of a reacting system as discrete and stochastic.
In Gillespie's Stochastic Simulation Algorithm~\cite{gillespie76}, the state of the
system is represented by a vector
$\textbf{x}=\textbf{X}(t)=(X_1(t),\cdots,X_N(t))$,
where $X_i(t)$ represents the number of $S_i$ individuals in \emph{V} at time $t$.
Gillespie assumed that for every reaction channel $R_j$, there is a constant $c_j$
such that $c_j dt$ is the average probability that a particular combination of reactant
individuals in $R_j$ will react accordingly in the next infinitesimal time interval $dt$.
To calculate the probability that a reaction $R_j$ will occur in $V$ in the next
infinitesimal time interval ($t,t+dt$), we must multiply $c_j dt$ by the total number
of distinct combinations of individuals in $V$ at time $t$ that are reactants
of $R_j$.
Let us denote such number by $h_j$(\textbf{x}).
Gillespie defines the \emph{propensity function}
$a_j$(\textbf{x}) for reaction $R_j$ as the product of $h_j$(\textbf{x})
and $c_j$, such that $a_j$(\textbf{x}) $dt$ is the probability that one
$R_j$ reaction will occur in the next infinitesimal time interval [$t,t+dt$).

Gillespie defines a Direct Method to implement his Stochastic Simulation
Algorithm~\cite{gillespie77}.
This version of Gillespie's SSA is defined as follows.
\begin{algo}\label{algo:SSA1}
Let $\{R_1,\ldots,R_M\}$ be a set of rewrite rules,
$X_1,\ldots,X_N$ be numbers of $N$ categories of individuals,
$maxtime$ be the time limit for the duration of the simulation.
\begin{description}
\item[Step 0]
Initialise simulation time $t$ to 0.
Compute propensity $a_i$ for every rewrite rule $R_i$.
\item[Step 1]
Compute the time increment $\tau$.
\item[Step 2]
Increase simulation time $t$ by time increment $\tau$. 
\item[Step 3]
If $t > maxtime$ then stop.
Otherwise select the next rule index $\mu$. 
\item[Step 4]
Execute rule $R_\mu$
and update numbers $X_1,\ldots,X_N$ of $N$ categories of
individuals and propensities $a_i$ for all rewrite rules $R_i$
affected by the application of $R_\mu$ accordingly.
Return to \textbf{Step 1}.
\end{description}
\end{algo}
Gillespie showed in his paper~\cite{gillespie77} that the time when next
reaction occurs (time $t+\tau$ calculated at Step 2 when reaction $R_\mu$
selected at Step 3 occurs)
is exponentially distributed with parameter
$a_0(\mathbf{x})=\sum_{i=1}^{M}a_i(\mathbf{x})$.
Gillespie used a general Monte Carlo method called \emph{inversion method}
to compute the exponentially distributed $\tau$ and $\mu$ from two uniformly
distributed random numbers as follows:
\begin{gather}
\tau = \frac{1}{a_0(\mathbf{x})}ln(\frac{1}{r_1})
\label{taucalc}\\
\mu=the\ integer\ for\ which\ \sum_{v=1}^{\mu-1} a_v(\mathbf{x}) <\
r_2a_0(\mathbf{x})\leq\ \sum_{v=1}^{\mu} a_v(\mathbf{x}) \label{muchoice}
\end{gather}
where $r_1, r_2$ are two real values uniformly distributed over interval [0,1] generated by a random number
generator.

In previous work Basuki, Cerone and Carvalho~\cite{cellcycleSpCLS}
extended Algorithm~\ref{algo:SSA1} to handle compartment selection.
This is useful when we have to simulate a biological system with multi-compartments
as is the case for molecular reactions occurring within cells.
Such a modified version of the Direct Method is described as follows.
\begin{algo}\label{algo:SSA2}
Let $\{R_1,\ldots,R_M\}$ be a set of $M$ rewrite rules,
$X_1,\ldots,X_N$ be numbers of $N$ categories of individuals,
$maxtime$ be the time limit for the duration of the simulation.
\begin{description}
\item[Step 0]
Initialise simulation time $t$ to 0.
Compute propensity $a_i$ for every rewrite rule $R_i$.
\item[Step 1]
Compute the time increment $\tau$.
\item[Step 2]
Increase simulation time $t$ by time increment $\tau$. 
\item[Step 3]
If $t > maxtime$ then stop.
Otherwise select the next rule index $\mu$ and the index $\theta$
of the compartment in which rule $R_\mu$ will occur. 
\item[Step 4]
Execute rule $R_\mu$ in the compartment with index $\theta$
and update numbers $X_1,\ldots,X_N$ of $N$ categories of
individuals and propensities $a_i$ for all rewrite rules $R_i$
affected by the application of $R_\mu$ accordingly.
Return to \textbf{Step 1}.
\end{description}
\end{algo}
Since reactions are confined within compartments, we need to extend Gillespie's 
algorithm to choose in which compartment reaction $R_{\mu}$ should occur.
Let $C$ be the number of compartments and $X_k^i$ the number of individuals
of kind $S_k$ in the $i$-th compartment. We define $X_k = \sum_{i=1}^C X_k^i$.

Let $a_j^i$ be the propensity of reaction $R_j$ occurring inside the $i$-th
compartment.
Then $a_j^i$ is defined as the product of $c_j$ by the number $h_j^i$ of distinct
combinations of reacting individuals of reaction $R_j$ within the $i$-th compartment.
We define  
$a_j = \sum_{i=1}^C a_j^i$.
If $t$ is the current simulaton time, then $t + \tau$ 
represents the time at which
next reaction occurs, with $\tau$ exponentially distributed with 
parameter $a_0=\sum_{j=1}^M a_j$.
Time increment $\tau$ is calculated as in Alghorithm~\ref{algo:SSA1}.
The index $\mu$ of the reaction that occurs at time $t + \tau$
and the index $\theta$ of the compartment in which such reaction occurs are calculated
as follows:
\begin{gather}
(\mu,\theta)=the\ integers\ f\! or\ which\ \sum_{j=1}^{\mu} \sum_{i=1}^{\theta-1} a_j^i <\
r_2a_0\leq\ \sum_{j=1}^{\mu} \sum_{i=1}^{\theta} a_j^i 
\end{gather}
where $r_2$ is a random real number which is uniformly distributed over interval [0,1].

\section{Extending Stochastic CLS}\label{sect:SCLSext}

The evolution of a system modelled by using Stochastic CLS is entirely
characterised by
the rewrite rules, which determine the occurrence of events in the system.
In this way the set of rewrite rules predicts all events that may occur.
This works well for biological systems, where all events are caused by
biochemical reactions which are governed by precise laws.

In ecological systems, instead, we need to deal with environmental
events, whose cause is often unknown or depends on a very complex
combination of factors, which are external to the system itself.
For example the dynamics of a population of a given species depends
not only on the interaction with other species within the same ecosystem,
such as predators, preys and competitors, but also on the occurrence of
environmental events such as climatic events (i.e.\ variation of
temperature and rainfalls) and events related to habitats (i.e.\ tree
clearing, desiccation of a water container, pollution, hunting and human
settlement).
Therefore, we assume the existence of a list of external events,
with information about the time when these events occur.
The occurrence of an external event may modify some environmental
information which affects the ecosystem evolution, such as temperature,
volume of water, desiccation, level of pollution.
Moreover, the list of external events may change dynamically.
For instance, an initial desiccation event for a water container
will be removed from the list after the occurrence of a rainfall event,
and will be replaced with a new desiccation event with a later desiccation time.

We extend Stochastic CLS by introducing a list $\extlist$ of external events.
The events in list $\extlist$ are sorted in increasing order based on the time
they are scheduled to occur.
When an external event occurs it updates information in the system state.
The updated information may be then used by rewrite rules.

In general, the environment is organised as several nested compartments, each
associated
with specific environmental information, which is relevant to the specific
ecosystem we
are modelling and may be modified by the occurrence of external events.
We further extend Stochastic CLS by attaching environmental information to the
looping operator.
This is similar to the extension of Stochastic CLS to Spatial CLS~\cite{SpatialCLS},
in which spatial information is added to the looping operator and sequence.

\begin{definition}\label{def:extsyntax}
\emph{Terms} $T$, \emph{Nonparallel Terms} $C$, \emph{Sequences} $S$ and
\emph{Environmental Information} $\envinf$ are given by the following grammar:
\begin{eqnarray*}
& T \qqop{::=} C^n \agr T \pipe T \ \ \ \ \ \ \ \ \ \ \ \ \ \ \! &
C \qqop{::=} S \agr \Loop{T}_\envinf \into T \\
& S \qqop{::=} \epsilon \agr a  \agr S \cdot S \ \ \ \ \ \ \ \ \ \ &
\envinf \qqop{::=} \lambda \agr a : V \agr \envinf \ \envinf
\end{eqnarray*}
\normalsize
where $a$ is a generic element of $\EE$, $\epsilon$ represents the empty sequence,
$\lambda$ represents the empty environmental information,
$V$ represents the information value
and $n \in \mathbb{N}$.
We denote with $\TT$, $\NPar$, $\Seq$ and $\EI$ the infinite set of terms,
nonparallel terms, sequences and environmental information, respectively.
\end{definition}
Note that in Definition~\ref{def:extsyntax} we have introduced a notation
to group identical nonparallel terms together.
For instance,
$C^5$ is equivalent to $C \pipe C \pipe C \pipe C \pipe C$.

Events in event list $\extlist$ update environmental information in the system state.
Every element of $\extlist$ is a triple $(N_\extlist,V_\extlist,t_\extlist)$, where
$N_\extlist$ is the name of the event,
$V_\extlist$ is a value that will be used to update the information field related to this event and
$t_\extlist$ is the time at which this event is scheduled to occur.
We assume the existence of an event handler algorithm which will handle the update of the term
representing the system state due to the occurrence of an event
$(N_\extlist,V_\extlist,t_\extlist)$.


To run the simulation using the extended version of Stochastic CLS, we need to modify the
version of Gillespie's Direct Method \cite{gillespie77} defined in
Algorithm~\ref{algo:SSA2}. 
In modelling population dynamics we  have to deal with the same problem we encounter
at cellular level: reactions occur in compartments.
Therefore, we extend the Direct Method for multi-compartments described in
Algorithm~\ref{algo:SSA2}
with additional steps to handle the execution of external events from the event list.
After computing the time of the next rewrite rule, we need to compare this time with the
time of the first event in the list, and execute the event with earlier occurrence time.
We propose the modified version of Direct Method as follows.

\begin{algo}\label{algo:SSA5}
Let $\{R_1,\ldots,R_M\}$ be a set of rewrite rules,
$X_1,\ldots,X_N$ be numbers of $N$ categories of organisms,
$\extlist$ be a list of events and
$maxtime$ be the time limit for the duration of the simulation.
\begin{description}
\item[Step 0]
Initialise simulation time $t$ to 0.
Compute propensity $a_i$ for every rewrite rule $R_i$.
\item[Step 1]
Compute the time increment $\tau$.
Let $(N_\extlist,V_\extlist,t_\extlist)$ be the first event from $\extlist$ with
$N_\extlist$ the name  of the event,
$V_\extlist$ the value needed to update the system state and
$t_\extlist$ the occurrence time of the event.
\item[Step 2]
If $t_\extlist < t + \tau$ then set $t$ to $t_\extlist$ and
then call the event handler algorithm to handle the new event
and return to \textbf{Step 1}.
Otherwise increase simulation time $t$ by time increment $\tau$. 
\item[Step 3]
If $t > maxtime$ then stop.
Otherwise select the next rule index $\mu$ and the index $\theta$
of the compartment in which rule $R_\mu$ will occur. 
\item[Step 4]
Execute rule $R_\mu$ in the compartment with index $\theta$
and update numbers $X_1,\ldots,X_N$ of $N$ categories of
organisms and propensities $a_i$ for all rewrite rules $R_i$
affected by the application of $R_\mu$ accordingly.
Return to \textbf{Step 1}.
\end{description}
\end{algo}
The event handler algorithm is specific to the external events
occurring in the system.
This algorithm updates system state and list of external events
and recomputes the propensities that have been affected by the change
of system state.

The simulation is affected by the propensity of every rewrite rule.
Propensity depends on the number of individuals in the population and the
rule rate constant.
External factors from the environment affect propensity values.
In general, we cannot associate a rule rate constant with each
rewrite rule, because the value of the rule rate depends on
environmental information, which changes according to
external events.
Since environmental information is incorporated in terms, to model
the rule rate we associate with that rule a function $f$
ranging over terms.

\begin{definition}\label{def:epattern} 
Let $TV$ be an infinite set of term
variables ranged over by $X,Y,Z,\ldots$,
$IV$ be an infinite set of information
variables ranged over by $x,y,z,\ldots$
and $NV$ be an infinite set of natural number
variables ranged over by $q,r,s,\ldots$\ 
Information Patterns $IP$, Term Patterns $TP$ and Patterns $P$ are defined as follows:
\begin{align*}
 IP \qqop{::=} I \agrette I \pipe x \quad\;
 CP \qqop{::=} S \agrette \Loop{T}_{IP} \into P \quad\;
 TP \qqop{::=} CP^q \agrette TP \pipe TP \quad\;
 P \qqop{::=} TP \agrette TP \pipe X
\end{align*}
where $X \in TV$ $x \in IV$ and $q \in NV$. We denote with $\PP$ the set of all patterns.
We denote with $Var(P)$ the set of variables in $P$.
\end{definition}

\begin{definition}\label{def:ext-instantiation}
The \emph{instantiation} is a partial function
$\sigma:TV \cup IV \cup NV \rightarrow \TT \cup \EI \cup \mathbb{N}$
such that $\sigma(TV) \subseteq \TT$, $\sigma(IV) \subseteq \EI$
and $\sigma(NV) \subseteq \mathbb{N}$.
We denote with $\Sigma$ the set of all possible term instantiations.
Given $P \in \PP$, we denote with $P\sigma$ the term obtained by replacing
all variables $X \in Var(P)$ with $\sigma(X)$.
\end{definition}

\begin{definition} A \emph{rewrite rule} is a 4-tuple ($f_c, P_L, P_R, f$),
usually written as
\[[f_c] \; P_L \stackrel{f}{\mapsto} P_R\]
where
$f_c : \Sigma \rightarrow \{ true , f\! alse \}$,
Var($P_R$) $\subseteq$ Var($P_L$), and $f : T \rightarrow \mathbb{R}^{\geq 0}$.
\end{definition}

The left pattern matches a portion of the term that models the system by using
an instantiation function $\sigma\in\Sigma$.
This portion of the system must also satisfy the constraint function $f_c$
to enable the rule to be applied.
A rate function $f$ associated with the rule will be applied to $P_L\sigma$.
After the rule is applied, $P_L\sigma$ is substituted by $P_R\sigma$.

\begin{definition}
An ecosystem is a triple $(T,\RR,\extlist)$, where $T$ is a term
representing the initial state of the system, $\RR$ is a set of rewrite
rules that represent the potential \emph{internal events} which may occur in the system,
and $\extlist$ is a list of \emph{external events}.
\end{definition}

\section{Modelling the Population Dynamics of {\it Aedes
albopictus}}\label{case:Aedes}

We use the formalism developed in the previous section to model
{\it Aedes albopictus} population dynamics.

\subsection{Modelling Information about a Mosquito}\label{sect:mosquitos}
We model each mosquito by using a looping and containment operator with
a parallel composition of symbols representing information about the mosquito
in the content part and a symbol $a$ in the loop part.
The information in the content part consists of the current development phase of
the mosquito and an indicator of whether the mosquito has sucked blood or not.
In our approach we only model females, assuming equal numbers of males and
females in the population.
In this way we do not need to model gender in the information of a mosquito. 

{\it Aedes albopictus} goes through 4 development phases in its life cycle:
egg, larva, pupa, and adult.
The larval stage is divided into 4 \emph{instars}~\cite{bradford05}.
The adult stage is divided into 8 \emph{gonotrophic cycles}~\cite{Delatte09}.
A gonotrophic cycle is a cycle in the adult life which consists of three phases
called Beklemishev phases~\cite{gonotrophicAnopheles}:
search for a host and blood-feeding,
digestion of the blood and egg maturation,
search for a suitable oviposition site and oviposition.
We use symbols $Egg$, $Larva$, $Pupa$ and $Adult$ to denote the 4 development
phases.
Since larva phase is divided into 4 instars, we use symbols $1$, ..., $4$ to
represent instars.
Analogously, we use symbols $1$, ..., $8$ to represent gonotrophic cycles.

An adult mosquito needs blood before ovipositing eggs.
We model this phenomenon by adding symbol $Blood$ to the content part of
the looping and containment operator defining the mosquito to represent an adult
mosquito that
has sucked blood.
The number of $Blood$ symbols in the content part indicates how many times that
mosquito has sucked blood.
For instance we represent 3 adult mosquitoes at gonotrophic cycle 1 that have sucked
blood twice and 5 larvae at instar 1 phase using the following term:\\
\( (\smallLoop{a}_\lambda \into (Adult\ |\ 1|\ Blood^2))^{3}\ |\
   (\smallLoop{a}_\lambda \into(Larva\ |\ 1))^{5} \).

\subsection{Modelling Compartments}\label{sect:compartments}
In Stochastic CLS compartments are modelled by using looping-containment
operators.
As we have seen in Definition~\ref{def:extsyntax} compartments play an important
role in our approach, because environmental information is attached to them.

{\it Aedes albopictus}, like other species of mosquitoes, spends
its immature stages
in water.
In particular, {\it Aedes albopictus} prefers to lay eggs
outdoors~\cite{SingaporeHabitat}.
Its natural breeding places are small, restricted, and shaded water collections
surrounded
by vegetation.
In urban areas, many man-made containers such as tin cans, pots, tires and
bottles are usually
stored outdoors and collect rainfall water, and thus become ideal breeding
places~\cite{spainalbopictus}.
Adult {\it Aedes albopictus} needs to suck blood before ovipositing.
However, {\it Aedes albopictus} only sucks blood during daytime.
Moreover, during immature stages, the duration of the stage is affected by
temperature
while death rate is affected by population density.
We can therefore define an outermost environmental compartment (we call it
\emph{environment}),
with the value of average daily temperature and daytime/nighttime as relevant
environmental
information, inside which there are several other compartments where immature
mosquitoes live
(we call them \emph{containers}).
Population density in one container is defined as the number of individuals
inside the
container divided by the water volume in the container.
Therefore, relevant environmental information for a container includes not only
temperature but also water volume and desiccation time.
Typical external events are sunrise and sunset, which determine switching
between daytime
and nighttime, temperature changes, which affect desiccation time by reducing
the volume
of water inside the containers and, as a result, increases population density,
and rainfalls, which increase the level of water in containers where mosquitoes
live, so
decreasing the population density.

Each kind of compartment has different environmental information.
The outermost compartment is the environment, to which we need to attach
information about current temperature and daylight.
Therefore, environment is modelled by a term
\[ \smallLoop{En}_{Temp:V_{Temp}\ Daylight:V_{Daylight}} \into (T)\]
where $V_{Temp}$ is a real number representing the current temperature,
$V_{Daylight}$ is a boolean representing whether it is daylight time and
$T$ is the term representing the population of {\it Aedes albopictus}. 

Immature {\it Aedes albopictus} live in small containers, modelled by
using
looping-containment operators with symbol $C$ inside the loop part.
For each container we attach the following environmental information:
\begin{itemize}
  \item an index to identify each container, to be used for container selection by
        Algorithm~\ref{algo:SSA5};
  \item the volume of water inside the container, to be used to compute population density;
  \item container temperature;
  \item three population density thresholds, to be used in the computation of death
    rates of mosquitoes living in the container;
  \item container desiccation time.
\end{itemize}
If $N_C$ is the number of containers in our model, we use natural numbers in $[1,N_C]$
to identify containers.
We model the volume of water in an abstract way by classifying containers as
$full$, $half\!-\!full$ and $empty$.
Population density thresholds, which are used to classify the population density in a
container and set the death rates accordingly will be defined in Section~\ref{sec:internal}.
Desiccation, or decrease of water, in a container is a process that depends on the
characteristic of the container.
A desiccation time, which measures how many days are needed to reduce the volume of water
in a container, is assigned to each container.
Container desiccation time will be defined in Section~\ref{sec:external}. 
As an example, term
\begin{align*}
Containers & \qqop{::=} \smallLoop{C}_{ind:1\ Temp:10\ Vol:empty\ \phi_1:100\ \phi_2:250\ \phi_3:300\ DTime:2.0} \into \epsilon\ | \\
& \smallLoop{C}_{ind:2\ Temp:10\ Vol:full\ \phi_1:50\ \phi_2:125\ \phi_3:150\ DTime:1.0} \into \epsilon
\end{align*}
defines two containers, one identified by number $1$, with no water, population density
thresholds 100, 250 and 300, and desiccation time 2 days, and one identified by number $2$,
full of water, with population density thresholds 50, 125 and 150, and desiccation time 1 day.

A population of immature and adult {\it Aedes albopictus} individuals is
modelled
as a parallel composition of looping and containment operators, each with symbol
$C$ inside the loop part to model a specific container and a parallel composition
of looping and containment operators (with symbol $a$ inside the loop part)
inside the content part to model the immature mosquitoes living inside that container,
and looping and containment operators with symbol $a$ inside the loop part to model
the adult {\it Aedes albopictus} individuals living in open space.
The whole population is then put inside another looping-containment operator with
symbol $En$ inside the loop part, which models the environment in which the
population lives.
In this way we model the environment in which a population lives as the outermost
compartment of the Stochastic CLS term that models the biological system of interest.

Given the two containers defined above, a daytime environment at a temperature of $10^\circ$ C
with a population of 8 adult mosquitoes at the first gonotrophic cycle, 5 of which have
sucked blood twice and 3 of which haven't sucked blood, and 2 empty containers is defined
as follows.
\begin{align*}
Pop & \qqop{::=} \smallLoop{En}_{Temp:10\ Daylight:true} \into (AdultPop\ |\ Containers)\\
AdultPop & \qqop{::=} (\smallLoop{a}_\lambda \into (
Adult | 1 | Blood^2))^{5}\ |\ \smallLoop{a}_\lambda \into (
Adult | 1))^{3}
\end{align*}
We assume that the temperature in all containers is the same as the temperature in the
environment.
Propagations of temperature changes in the environment to the containers are handled by
the event handler algorithm as we will explain in Section~\ref{sec:external}.

\subsection{Modelling Internal Events}\label{sec:internal}
We have seen in Section~\ref{sect:mosquitos} that the lifecycle of {\it Aedes albopictus}
consists of the following 14 stages: egg, larva (instar 1--4), pupa and
adult (8 gonotrophic cycles).
Internal events describe transitions between some of these stages as well
as other events occurring at a specific stage.
We identify 29 internal events and we model the effect of each of them
on the system by a rewrite rule:
\begin{description}
	\item[Rule R1] egg hatch
	\item[Rules R2--R4] transitions between instars
	\item[Rule R5] pupation
	\item[Rule R6] adult emergence
	\item[Rule R7] blood sucking
	\item[Rules R8--R15] oviposition at each gonotrophic cycle
	\item[Rules R16--29] death at each stage of the life cycle (14 events)
\end{description}

Rules R1--R5, which model transitions between immature development stages,
rule R6, which models the transition from the last immature development
stage to the first adult stage,
and rules R16--R21, which model the death events in such stages,
are shown in
\begin{figure}
\begin{center}
\framebox{
\begin{minipage}{0.9\textwidth}
\small
\begin{align*}
\smallLoop{C}_x \into \big(Y | \smallLoop{a}_\lambda \into (Egg | X)\big) \widesrewrites{f_1} & \smallLoop{C}_x \into \big(Y | \smallLoop{a}_\lambda \into (Larva | 1 | X)\big) &
\mbox{(R1)}\\
\smallLoop{C}_x \into \big(Y | \smallLoop{a}_\lambda \into (Larva | 1 | X)\big) \widesrewrites{f_2} & \smallLoop{C}_x \big(Y | \smallLoop{a}_\lambda \into (Larva | 2 | X)\big) & \mbox{(R2)}\\
\smallLoop{C}_x \into \big(Y | \smallLoop{a}_\lambda \into (Larva | 2 | X)\big) \widesrewrites{f_3} & \smallLoop{C}_x \into \big(Y | \smallLoop{a}_\lambda \into (Larva | 3 | X)\big) & \mbox{(R3)}\\
\smallLoop{C}_x \into \big(Y | \smallLoop{a}_\lambda \into (Larva | 3 | X)\big) \widesrewrites{f_4} & \smallLoop{C}_x \into \big(Y | \smallLoop{a}_\lambda \into (Larva | 4 | X)\big) & \mbox{(R4)}\\
\smallLoop{C}_x \into \big(Y | \smallLoop{a}_\lambda \into (Larva | 4 | X)\big) \widesrewrites{f_5} & \smallLoop{C}_x \into \big(Y | \smallLoop{a}_\lambda \into (Pupa | X)\big) & \mbox{(R5)}\\
\smallLoop{C}_x \into \big(Y | \smallLoop{a}_\lambda \into (Pupa | X)\big) \widesrewrites{f_6} & \smallLoop{C}_x \into Y\ \mid\  \smallLoop{a}_\lambda \into (Adult | 1 | X) & \mbox{(R6)}\\
\smallLoop{C}_x \into \big(\smallLoop{a}_\lambda \into(Egg | X)\mid Y\big) \widesrewrites{f_{16}} & \smallLoop{C}_x \into Y & \mbox{(R16)}\\
\smallLoop{C}_x \into \big(\smallLoop{a}_\lambda \into (Larva | 1 | X)\mid Y\big) \widesrewrites{f_{17}} &  \smallLoop{C}_x \into Y & \mbox{(R17)}\\
\smallLoop{C}_x \into \big(\smallLoop{a}_\lambda \into(Larva | 2 | X)\mid Y\big) \widesrewrites{f_{18}} & \smallLoop{C}_x \into Y & \mbox{(R18)}\\
\smallLoop{C}_x \into \big(\smallLoop{a}_\lambda \into (Larva | 3 | X)\mid Y\big) \widesrewrites{f_{19}} & \smallLoop{C}_x \into Y & \mbox{(R19)}\\
\smallLoop{C}_x \into \big(\smallLoop{a}_\lambda \into (Larva | 4 | X)\mid Y\big) \widesrewrites{f_{20}} & \smallLoop{C}_x \into Y & \mbox{(R20)}\\
\smallLoop{C}_x \into \big(\smallLoop{a}_\lambda \into (Pupa | X)\mid
Y\big) \widesrewrites{f_{21}} & \smallLoop{C}_x \into Y & \mbox{(R21)}
\end{align*}
\normalsize
\end{minipage}
}
\end{center}
\caption{Rewrite rules for the immature stages of {\it Aedes
albopictus}}\label{fig:immaturerules}
\end{figure}
Figure~\ref{fig:immaturerules}.

The duration of an immature stage depends on temperature and is measured in degree-days.
Degree-days for each immature stage is defined as the number of days it takes for an individual
in that stage to develop at $1^\circ$C above the minimum temperature for development
(MTD)~\cite{HawaiiCulex}.
Following this definition, we define the values of temperature in the environmental information
as the difference between the actual temperature and MTD.

If $d_i$ is the average duration of the $i$-th development stage, then the rate constant
of the rule modelling the transition from stage $i$ to the next stage is $1/d_i$.
This is true if there are no other events occurring during this stage.
For every immature development stage we define
one rule for the transition to the next stage and another one for the death
event.
Rate functions for transitions in immature stages and death events are computed
by multiplying $1/d_i$ by
survivability rate at $i$-th stage and by death rate at $i$-th stage,
respectively.
We assume that the sum of death rate and survivability rate at one development
stage is
equal to one.
Since the duration of an immature stage depends on temperature, the rate is then multiplied
by the difference between the current temperature and MTD.
The death rate at an immature stage is defined locally for each container and depends on the
population density of the container.
We classify the population density in a container into 4 classes of density: sparse, normal,
crowded and overcrowded.
We define 3 thresholds to be used to classify density: $\phi_1, \phi_2$ and $\phi_3$. 
These three thresholds are part of the environmental information attached to each container.
The rate functions for rules R1--R6 and R16--R21 are computed as follows:
\begin{equation}
f_i(\smallLoop{C}_\envinf \into \big(T\big)) = \left\{ \begin{array}{ll} \frac{V_{Temp}\cdot (1 - DR(i,n,V_{Vol},V_{\phi_1},V_{\phi_2},V_{\phi_3}))}{DD(i)} & \mbox{if $i \in [1,6]$} \medskip\\
\frac{V_{Temp}\cdot DR(i-15,n,V_{Vol},V_{\phi_1},V_{\phi_2},V_{\phi_3})}{DD(i-15)} & \mbox{if $i \in [16,21]$} \end{array} \right.
\end{equation}
where
\begin{itemize}
  \item $i$ is the index of the rewrite rule,
	\item $\envinf = ind\!\!:\!\!k\ \ Temp\!\!:\!\!V_{Temp}\ \ Vol\!\!:\!\!V_{Vol}\ \ \phi_1\!\!:\!\!V_{\phi_1}\ \ \phi_2\!\!:\!\!V_{\phi_2}\ \ \phi_3\!\!:\!\!V_{\phi_3}\ \ DTime\!\!:\!\!V_{DTime}$ 
is the environmental information attached to the container to which rule $Ri$ is applied,
  \item $V_{Temp}$ is the container temperature, 
  \item $DR(j,n,V_{Vol},V_{\phi_1},V_{\phi_2},V_{\phi_3})$ is the death rate
function at immature stage $j$ for the container which contains $n$ immature
mosquitoes, with density thresholds $\phi_1,\phi_2,\phi_3$ and contains a volume
$V_{Vol}$ of water,
  \item $DD(j)$ represents the duration of stage $j$ in degree-days.
\end{itemize}

We use 4 classes of population density (sparse, normal, crowded and overcrowded)
to define death rate in our model.
We use the following assumptions for all containers:
\begin{itemize}
\item threshold values used to classify population density in a container are defined
  for the case in which the container is full of water,
\item the baseline death rate of stage $i$ is the death rate of the population
  in a container whose population density is normal,
\item when population in one container is overcrowded or there is no more water
  in the container only death events can occur, so the death rate is set to 1,
\item death rate increases by 20\% above the baseline death rate if population
  density is crowded,
\item death rate decreases by 20\% below the baseline death rate if population
  density is sparse,
\item when a container is only half full, the values of thresholds used to classify
  the population density are divided by 2.
\end{itemize}
We define the death rate function
\( DR : \mathbb{N} \times \mathbb{N} \times \EE\times \mathbb{N} \times \mathbb{N}
                   \times \mathbb{N} \rightarrow \mathbb{R} \)
as follows:
\small
\begin{equation}
DR(j,n,V,\phi_1,\phi_2,\phi_3) = \left\{ \begin{array}{ll} 
1 & \mbox{if $V$ is $empty$} \\
1 & \mbox{if $V$ is $full$ and $n \geq \phi_3$} \\
1.2 \cdot BDR(j) & \mbox{if $V$ is $full$ and $\phi_2 \leq n < \phi_3$} \\
BDR(j) & \mbox{if $V$ is $full$ and $\phi_1 \leq n < \phi_2$} \\
0.8 \cdot BDR(j) & \mbox{if $V$ is $full$ and $n < \phi_1$} \\
DR(j,2n,full,\phi_1,\phi_2,\phi_3) & \mbox{if $V$ is $half\!-\!full$}
                                         \end{array} \right.
\end{equation}
\normalsize
where $BDR(j)$ is the baseline death rate for phase $j$ of the life cycle,
$n$ is the number of immature mosquitoes in the container,
$\phi_1,\phi_2,\phi_3$ are the container density thresholds and
$V$ is the volume of water in the container.

The adult life of an {\it Aedes albopictus} is divided into 8 gonotrophic
cycles.
Every gonotrophic cycle consists of two internal events: blood sucking and oviposition.
The oviposition (the sixth internal event) is also a transition from one gonotrophic
cycle to the next gonotrophic cycle.
Figure~\ref{fig:rules1}
\begin{figure}[t]
\begin{center}
\framebox{
\begin{minipage}{0.9\textwidth}
\small
\begin{align*}
& \smallLoop{En}_{Daylight:true \ x} \into (Y | \smallLoop{a}_{\lambda} \into (Adult | X | Blood^q)) \widesrewrites{f_7} & \\ 
& \smallLoop{En}_{Daylight:true \ x} \into (Y | \smallLoop{a}_{\lambda} \into (Adult | X | Blood^{q+1})) & \mbox{(R7)} \\
[q > \varphi] \; & \smallLoop{En}_x \into (Y |\smallLoop{a}_\lambda \into (Adult | 1 | X | Blood^q) | \smallLoop{C}_{y} \into Z) \widesrewrites{f_8} \smallLoop{En}_x \into & \\
& (Y |\smallLoop{a}_\lambda \into (Adult | 2 | X) |\smallLoop{C}_{y} \into (Z | (\smallLoop{a}_\lambda \into (Egg | X))^{eggs(1)}) & \mbox{(R8)}\\
[q > \varphi] \; & \smallLoop{En}_x \into (Y |\smallLoop{a}_\lambda \into (Adult | 2 | X | Blood^q) | \smallLoop{C}_{y} \into Z) \widesrewrites{f_{9}} \smallLoop{En}_x \into & \\
& (Y |\smallLoop{a}_\lambda \into (Adult | 3 | X) |\smallLoop{C}_{y} \into (Z | (\smallLoop{a}_\lambda \into (Egg | X))^{eggs(2)}) & \mbox{(R9)}\\
[q > \varphi] \; & \smallLoop{En}_x \into (Y |\smallLoop{a}_\lambda \into (Adult | 3 | X | Blood^q) | \smallLoop{C}_{y} \into Z) \widesrewrites{f_{10}} \smallLoop{En}_x \into & \\
& (Y |\smallLoop{a}_\lambda \into (Adult | 4 | X) |\smallLoop{C}_{y} \into (Z | (\smallLoop{a}_\lambda \into (Egg | X))^{eggs(3)}) & \mbox{(R10)}\\
[q > \varphi] \; & \smallLoop{En}_x \into (Y |\smallLoop{a}_\lambda \into (Adult | 4 | X | Blood^q) | \smallLoop{C}_{y} \into Z) \widesrewrites{f_{11}} \smallLoop{En}_x \into & \\
& (Y |\smallLoop{a}_\lambda \into (Adult | 5 | X) |\smallLoop{C}_{y} \into (Z | (\smallLoop{a}_\lambda \into (Egg | X))^{eggs(4)}) & \mbox{(R11)}\\
[q > \varphi] \; & \smallLoop{En}_x \into (Y |\smallLoop{a}_\lambda \into (Adult | 5 | X | Blood^q) | \smallLoop{C}_{y} \into Z) \widesrewrites{f_{12}} \smallLoop{En}_x \into & \\
& (Y |\smallLoop{a}_\lambda \into (Adult | 6 | X) |\smallLoop{C}_{y} \into (Z | (\smallLoop{a}_\lambda \into (Egg | X))^{eggs(5)}) & \mbox{(R12)}\\
[q > \varphi] \; & \smallLoop{En}_x \into (Y |\smallLoop{a}_\lambda \into (Adult | 6 | X | Blood^q) | \smallLoop{C}_{y} \into Z) \widesrewrites{f_{13}} \smallLoop{En}_x \into & \\
& (Y |\smallLoop{a}_\lambda \into (Adult | 7 | X) |\smallLoop{C}_{y} \into (Z | (\smallLoop{a}_\lambda \into (Egg | X))^{eggs(6)}) & \mbox{(R13)}\\
[q > \varphi] \; & \smallLoop{En}_x \into (Y |\smallLoop{a}_\lambda \into (Adult | 7 | X | Blood^q) | \smallLoop{C}_{y} \into Z) \widesrewrites{f_{14}} \smallLoop{En}_x \into & \\
& (Y |\smallLoop{a}_\lambda \into (Adult | 8 | X) |\smallLoop{C}_{y} \into (Z | (\smallLoop{a}_\lambda \into (Egg | X))^{eggs(7)}) & \mbox{(R14)}\\
[q > \varphi] \; & \smallLoop{En}_x \into (Y |\smallLoop{a}_\lambda \into (Adult | 8 | X | Blood^q) | \smallLoop{C}_{y} \into Z) \widesrewrites{f_{15}} & \\
& \smallLoop{En}_x \into (Y |\smallLoop{C}_{y} \into (Z |
(\smallLoop{a}_\lambda \into (Egg | X))^{eggs(8)}) & \mbox{(R15)}
\end{align*}
\normalsize
\end{minipage}
}
\end{center}
\caption{Rewrite rules for blood-sucking and oviposition events of {\it Aedes
albopictus}}\label{fig:rules1}
\end{figure}
shows the rewrite rules modelling blood-sucking and oviposition events.
Rule $R7$ models the blood sucking by adult mosquitoes.
All adult mosquitoes have the same probability of sucking blood.
We assume that a mosquito always sucks a constant amount of blood.
To oviposit, the amount of blood sucked by an adult female must be above a threshold
(represented by $\varphi$ in rules R8--R15).

Rules R8--R15 model the oviposition for the 8 gonotrophic cycles of the
mosquito.
We assume that all adults die after ovipositing at the 8th gonotrophic cycle.
The number of eggs any female can produce in each gonotrophic cycle is between
45 and 80. This number declines over age.
We model this by defining function $eggs$, for each gonotrophic cycle $j$
of the mosquito.
\begin{equation}
eggs(j) = \left\{ \begin{array}{ll} 
40 \quad \mbox{if $j = 1$}  \qquad \qquad 30 \quad \mbox{if $j = 5$} \\
37 \quad \mbox{if $j = 2$}  \qquad \qquad 27 \quad \mbox{if $j = 6$} \\
35 \quad \mbox{if $j = 3$}  \qquad \qquad 25 \quad \mbox{if $j = 7$} \\
32 \quad \mbox{if $j = 4$}  \qquad \qquad 22 \quad \mbox{if $j = 8$} \\
\end{array} \right.
\end{equation}
Although the number of eggs produced by a female mosquito at the $j$-th
gonotrophic cycle is between 45 and 80, $eggs(j)$ only returns half of this
value to take into account that we only model female individuals.

\begin{figure}
\begin{center}
\framebox{
\begin{minipage}{0.9\textwidth}
\small
\begin{align*}
\smallLoop{En}_x \into (Y |\smallLoop{a}_\lambda \into (Adult | 1 | X)) \widesrewrites{f_{22}} & \smallLoop{En}_x \into Y & \mbox{(R22)}\\
\smallLoop{En}_x \into (Y |\smallLoop{a}_\lambda \into (Adult | 2 | X)) \widesrewrites{f_{23}} & \smallLoop{En}_x \into Y & \mbox{(R23)}\\
\smallLoop{En}_x \into (Y |\smallLoop{a}_\lambda \into (Adult | 3 | X)) \widesrewrites{f_{24}} & \smallLoop{En}_x \into Y & \mbox{(R24)}\\
\smallLoop{En}_x \into (Y |\smallLoop{a}_\lambda \into (Adult | 4 | X)) \widesrewrites{f_{25}} & \smallLoop{En}_x \into Y & \mbox{(R25)}\\
\smallLoop{En}_x \into (Y |\smallLoop{a}_\lambda \into (Adult | 5 | X)) \widesrewrites{f_{26}} & \smallLoop{En}_x \into Y & \mbox{(R26)}\\
\smallLoop{En}_x \into (Y |\smallLoop{a}_\lambda \into (Adult | 6 | X)) \widesrewrites{f_{27}} & \smallLoop{En}_x \into Y & \mbox{(R27)}\\
\smallLoop{En}_x \into (Y |\smallLoop{a}_\lambda \into (Adult | 7 | X)) \widesrewrites{f_{28}} & \smallLoop{En}_x \into Y & \mbox{(R28)}\\
\smallLoop{En}_x \into (Y |\smallLoop{a}_\lambda \into (Adult | 8 | X))
\widesrewrites{f_{29}} & \smallLoop{En}_x \into Y & \mbox{(R29)}
\end{align*}
\normalsize
\end{minipage}
}
\end{center}
\caption{Rewrite rules for death events in adult phases of {\it Aedes
albopictus} life cycle}\label{fig:rules2}
\end{figure}

Finally figure \ref{fig:rules2} shows rules R22--R29,
which model the death at every adult stage.
Rule rates for rules R7--R15 and R22--R29 are defined as follows:
\begin{equation}
 f_i = \left\{ \begin{array}{ll} \frac{1}{d(i)} & \mbox{if $i = 7$} \medskip \\
\frac{(1 - BDR(i))}{d(i)} & \mbox{if $i \in [8,15]$} \medskip \\
\frac{BDR(i-14)}{d(i-14)} & \mbox{if $i \in [22,29]$} \end{array} \right. \end{equation}
where $d(i)$ is the duration of stage $i$ and $BDR(i)$ is the death rate at stage $i$.

All rules presented in this section are implemented by using Maude rewrite laws.

\subsubsection{Implementation Strategies}\label{sec:implint}

Since we use Maude to implement our model, which mosquito is chosen
in the application of rule R7 depends on the strategy implemented in Maude.
To guarantee fairness we implement our own strategy in choosing the mosquito
with the smallest number of blood sucking times first.

All adult mosquitoes in a given development stage that have sucked enough
blood have the same probability of ovipositing.
Therefore we consider one rule for each development stage (rules R8--R15).
We have to deal with the same problem (of choosing the mosquito to oviposit)
as in rule R7.
To guarantee fairness we define a strategy of choosing the mosquito, based on
how many times the mosquito has sucked blood.
As a consequence of the strategy defined for rule R7 the number of times a
mosquito sucks blood is proportional to the time spent in adult stages.
Our strategy will choose the mosquito with the biggest number of blood sucking
times of ovipositing first.

We also implement a strategy for choosing the container in which a
mosquito oviposits.
This strategy randomly chooses the container in which the mosquito oviposits.

The three strategies we have defined in this section have a different purpose
from the strategy defined by Basuki, Cerone and Milazzo \cite{SCLSMaude},
which was used to choose which rewrite rule to apply during a simulation.
Instead, the strategies defined in this section are used to choose which portion
of the term that models the system state matches the lefthand side of a rewrite rule.

\subsection{Modelling External Events}\label{sec:external}
External events are events that cannot be controlled by the system.
These events are usually used to model changes
in the environment that affect the population.
Every event is modelled as a triplet $(N,V_,t)$, where
the event name $N$ is used to distinguish the kind of event,
the event value $V$ is used to update the environmental information in the system
state and
the event time $t$ is the time when the event is scheduled to occur.
Event names and values will be explained in the next paragraphs.
Event time $t$ is a non-negative real number and measures time in days.
The integer part of $t$ represents the day and the fractional part represents
the time of the day at which an event should occur.
For instance $t = 1.5$ means that the event is scheduled to occur on day 1 at 12 pm,
and $t = 4.125$ means that the event is scheduled to occur on day 4 at 3 am.


As explained in Section~\ref{sect:compartments}, for each container
there are seven kinds of environmental information in our model:
container index, container temperature, volume of water in the container,
three container thresholds for population density and container desiccation
time.
External events must deal with these kinds of environmental information.
We define four kinds of event:
light change event, change of temperature, desiccation, and rainfall.
Light change events are scheduled twice a day, one at sunrise and another at
sunset.

A sunrise event changes the $Daylight$ information associated with the
environment from $false$ to $true$.
A sunset event changes the $Daylight$ information from $true$ to $false$.
A change of temperature event updates temperature in all compartments.
A desiccation event updates the volume of water in a specific container.
A rainfall event updates the volume of water in all containers.
Container indices are used by the event handler algorithm to handle all
events that occur.
Population density thresholds are used to compute propensity after population
density in one container is updated due to the occurrence of a desiccation or
a rainfall event.
Container desiccation time is used to schedule new desiccation events due to
the occurrence of a desiccation or a rainfall event.

A light change event is modelled as a triplet $(Light,V,t)$.
The time when the sun rises and the time when the sun sets depend on the position
of a place on the earth and the time of the year.
Value $V$ determines whether the event is sunrise ($V = sunrise$)
or sunset ($V = sunset$) event.
For instance in a place where in a winter day the sun rises at 8 am and sets at 5 pm,
the sunrise event on day 1 is modelled as a triplet
$(Light,sunrise,1.33)$
and the sunset event on the same day is modelled as
$(Light,sunset,1.71)$.

Temperature affects the duration of immature phases of the mosquito development.
We model a temperature change as a triplet $(Temp,V_{Temp},t)$, which is interpreted
as the event of setting the temperature to a new value $V_{Temp}$ starting from
time $t$.
We consider only the average daily temperature.
We schedule one temperature change event every day at midnight.
So a triplet $(Temp,10,3.0)$ means that the average temperature on day 3
is $10^\circ$C above the MTD of {\it Aedes albopictus}.

The desiccation event is modelled as a triplet $(Desic,i,t)$ which is
interpreted as a desiccation in the container with index $i$ at time $t$.
We assume that the desiccation time depends on container type and measure
this time as the number of days needed to reduce the water volume by one level
(from $full$ to $half\!\!-\!\!full$ or from $half\!\!-\!\!full$ to $empty$).
Initially, we introduce one event for each container in list $L$ scheduled
according to the desiccation time of the container to which it refers.
Every time a desiccation event occurs and the container is not yet empty,
another desiccation event is scheduled to reduce the water volume to the next level.
For instance, if the system state is represented as:
\[\smallLoop{En}_\envinf \into (\smallLoop{C}_{ind:1\ Vol:empty\ DTime:2.0\ \envinf'}
            \into T'\ | \smallLoop{C}_{ind:2\ Vol:full\ DTime:1.5\ \envinf''} \into T'')\]
where $\envinf, \envinf'$ and $\envinf''$ represent part of environmental information which is not
relevant for desiccation events, $T'$, $T''$ are terms representing population of Aedes
albopictus inside container 1 and 2, respectively, and the first event in list $\extlist$ is
$(Desic,2,1.0)$, then at time 1.0 the system state becomes
\[\smallLoop{En}_\envinf \into (\smallLoop{C}_{ind:1\ Vol:empty\ DTime:2.0\ \envinf'}
            \into T'\ | \smallLoop{C}_{ind:2\ Vol:half-full\ DTime:1.5\
\envinf''} \into T'')\, .\]
The event $(Desic,2,1.0)$ is removed from and a new desiccation event
$(Desic,2,2.5)$ is added to list $\extlist$.

In our model we only consider containers stored outdoors.
In this way, rainfalls are scheduled events that increase the water
volume level in all containers.
Rainfalls are assumed to be prescheduled initially.
Every time a rainfall event occurs, all desiccation events have to be removed
from the list and new desiccation events should be added.
We classify rainfalls as $heavy$ and $light$.
A heavy rainfall increases the water volume level of all containers to $full$.
A light rainfall increases the water volume level of all containers from $empty$
to $half\!\!-\!\!full$ or from $half\!\!-\!\!full$ to $full$.
The rainfall event is modelled  as a triplet $(Rain,lev,t)$ which represents
a rainfall event with level $lev$ ($heavy$ or $light$) starting at time $t$.
For instance, if the system state is represented as:
\[\smallLoop{En}_\envinf \into (\smallLoop{C}_{ind:1\ Vol:empty\ DTime:2.0\ \envinf'}
    \into T'\ | \smallLoop{C}_{ind:2\ Vol:half-full\ DTime:1.5\ \envinf''} \into T'')\]
and list $\extlist$ contains three events $(Rain,light,1.25)$, $(Desic,2,1.5)$ and
$(Desic,1,2.0)$,
then at time 1.25 the system state becomes
\[\smallLoop{En}_\envinf \into (\smallLoop{C}_{ind:1\ Vol:half-full\ DTime:2.0\ \envinf'}
    \into T'\ | \smallLoop{C}_{ind:2\ Vol:full\ DTime:1.5\ \envinf''} \into T'')\]

The three events are removed from the list and two new desiccation events
$(Desic,2,2.75)$ and $(Desic,1,3.25)$ are added to the list.

The event handler algorithm is very simple.
Given a list $\extlist$ of events,
$N_C$ containers and
a term $T$ that represents the system state,
the algorithm removes the first event
$(N_\extlist,V_\extlist,t_\extlist)$
from $\extlist$ and performs the different actions decribed above
according to the value of
$N_\extlist \in \{ Light , Temp , Desic , Rain \}$.
The removal of the first event from the list and the subsequent
actions are implemented by using Maude rewrite laws.

\subsection{In silico Experiment and Analysis}
As already mentioned, we have implemented our model in Maude.
We have then run a simulation by using data collected during
May--November 2009 in the province of Massa-Carrara (Tuscany, Italy)
in 11 $CO_2$ mosquito traps.
The 11 traps have captured a total of 3535 {\it Aedes albopictus}
individuals, and have been checked at the following dates:
8 May (4 Aedes a.),
15 May (25 Aedes a.),   
19 May (81 Aedes a.),	   
5 June (33 Aedes a.),	   
18 June (167 Aedes a.),   
3 July (360 Aedes a.),   
14 July (561 Aedes a.),   
29 July (381 Aedes a.),   
19 August (486 Aedes a.),   
3 September (471 Aedes a.),	   
19 September (276 Aedes a.),   
23 September (292 Aedes a.),  
14 October (398 Aedes a.).  
Note that traps need to be charged with $CO_2$ in order to work, and that the
charge allows the trap to work for one day. Hence, data refer to captures of
mosquitoes in one day for each considered date. This way of sampling mosquito
populations follows standard practice.

\begin{figure}[tb]
    \begin{center}
       \includegraphics[scale=0.65]{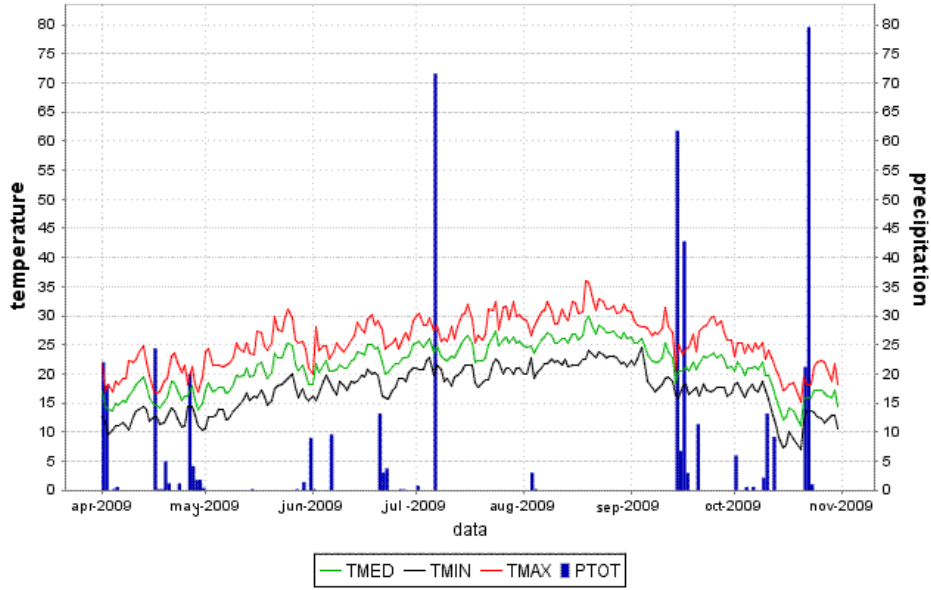}
        \caption{Temperature and Rainfall in Massa Carrara, Italy}
       \label{fig:data}
    \end{center}
\end{figure}

Figure~\ref{fig:data} shows the climatic data (temperature in \textdegree C and
rainfalls in $mm$) during May--November 2009 in Massa-Carrara province.
In our simulation we use  $8.8^{\circ}$C as MTD~\cite{TengApperson} and
11 containers.
Each container has carrying capacity of 100--250 organisms and 
desiccation time between 4.5 and 9.0 days.

In our simulation we initialise the population with 4 adult mosquitoes
(which equals the number of adult mosquitoes collected on 8 May 2009)
and 10 immature mosquitoes in each of the 11 containers,
6 eggs, 2 instar-1 larvae, 1 instar-2 larva and 1 instar-3 larva.
The water volume level in each container is initially set to half-full.
We also set initial desiccation events according the desiccation times
of the containers.
Let $t_0$ be the time when the simulation starts and $DT_i$ be the
desiccation time of container identified by index $i$, then
we set initial desiccation events at time $t_0 + DT_i$ for
$i = 1$ to $11$.

\begin{figure}[tb]
    \begin{center}
       \includegraphics[width=8.9cm]{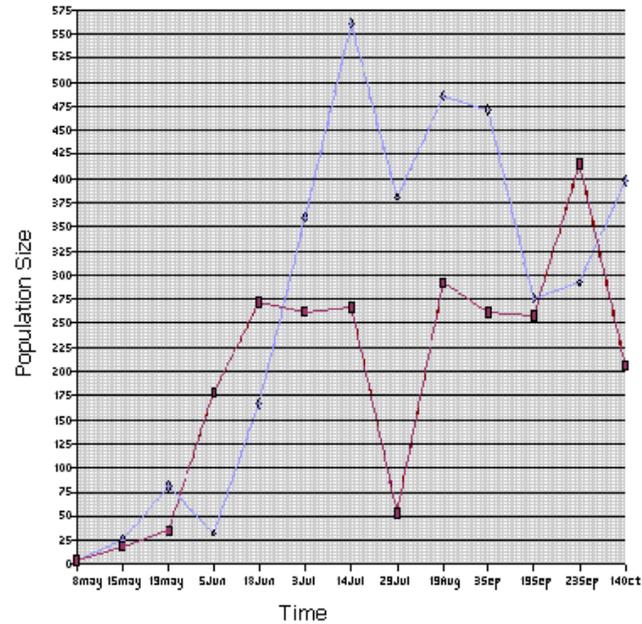}
        \caption{Comparison of in silico simulation (dark line) with data sampled from mosquito traps (light line)}
       \label{fig:compare}
    \end{center}
\end{figure}

Figure \ref{fig:compare} shows the result of our simulation
compared with the population sampling produced by using the 11 traps.
We can notice some differences between the simulation results and the
field sampling. For
example, the number of mosquitoes in the sampling decreases between 19 May
and 5 June, whereas in the simulation such number rapidly increases.
This probably happens because of the coarse classification of rainfalls
in our model: a very tiny rainfall, with neglectable effect in reality,
which occurs just before 19 May, is classified as light rain and, as a
result, increases the level of water of the containers in the simulation.
This may indicate that we need to improve our model by using
a finer classification of rainfalls.

The number of mosquitoes captured in traps rapidly increases from 18 June
to 14 July, probably due to rainfalls.
However, no population growth is shown by the simulation during that period.
This may be due to an overweighed effect that temperature decrease has in
our model on immature stage duration and death rates.
It may also be due to too small values for dessication times used in our model.

In the simulation the effect of the heavy rainfalls that occur just before
19 September immediately causes a population increase on 23 September,
but the subsequent decrease in rainfalls and heavy decrease in temperature
lead to a population decrease on 14 October.
In the field sampling, instead, population samples keep increasing from 19 September
to 14 October.
This difference might indicate that decrease of rainfalls and temperature take a longer
time in reality to affect the population growth than in our simulation.
This might be again due to too small values for dessication times used in our model.
Moreover, a decrease in temperature might cause a slower desiccation, a phenomenon
that is not considered in our model.

\section{Conclusions}

We have presented an extension of the Calculus of Looping Sequences aimed at describing population dynamics and ecosystems.
The extension consists in allowing a list of external events to be provided by the modeller in order to describe environmental events such as changes in the climatic conditions. The modeller has also to provide an event handler algorithm that is used by the simulation algorithm associated with the extended formalism. The event handler algorithm is invoked every time an external event is planned to occur and it changes the simulation state in accordance with the type of the considered event.

We have used the extended formalism to give a model of a population of {\em Aedes albopictus}, an aggressive mosquito that is well known as an important disease vector. A simulator for this model has been developed via translation into the specification language Maude. We have compared results of simulations of our model with real data obtained from the sampling of mosquitoes during May-November 2009 in the province of Massa-Carrara (Tuscany, Italy). Since changes in the temperature and rainfalls have a significant effect on the mosquito population dynamics, we have exploited data on such environmental events (in the same area and the same period of the sampling of mosquitoes) to construct a list of external events for the model. 

The results of our simulations show some differences from the real data. However, these differences seem to be motivated by some restrictive modelling choices that could be revised in order to construct an improved and finer model.
Improvements to the model are hence part of our future work, which includes also
\begin{itemize}
  \item modelling of populations of other disease vector mosquitos such as {\em Aedes aegypti};
  \item study of dynamics of populations in other geographic areas;
  \item study of different control policies to the mosquito population.
\end{itemize}
It would be particularly interesting to study the effects of events such as periodic cleaning of containers and use of pesticides on the mosquito population to choose the most promising control policy. Such a policy could then be experimented in the field, and the results obtained could be used to further validate the model. A method to choose the best mosquito population control policy would be of interest in particular in those areas in which such mosquitoes act as vectors of diseases.

\section*{Acknowledgment}
Antonio Cerone would like to thank Syed Mohamed Aljunid and Jamal Hisham Hashim
for their hospitality and helpful discussion at UNU-IIGH in Kuala Lumpur and
for further email discussions which inspired this work.

This work has been supported partly by UNU-IIST core funding, partly by
a grant from Fondazione Monte dei Paschi di Siena for the project
``Un approccio etologico/computazionale allo studio dei meccanismi dell'evoluzione
delle specie animali'' and partly by a grant for international inter-university
collaborations from the Italian Ministry for Education and Scientific Research
(MIUR).
\bibliography{aedes}
\bibliographystyle{eptcs}

\end{document}